\documentclass[aps,twocolumn,superscriptaddress,prl,amsmath]{revtex4-2}
\usepackage{multirow}
\usepackage[latin9]{inputenc}
\usepackage{epstopdf}
\usepackage{mathrsfs}
\usepackage{txfonts}
\usepackage{amssymb}
\usepackage{graphicx,subfigure,float}
\usepackage{dcolumn}
\usepackage{bbm}
\usepackage{bm}
\usepackage{color}
\usepackage[colorlinks, linkcolor=blue, citecolor=blue, urlcolor=blue]{hyperref}
\usepackage{lipsum}
\usepackage{gensymb}

\begin{document}
\title{Emergent Surface Altermagnetism}
\author{Yuzhong Hu}
 \affiliation{Key Laboratory of Low Dimensional Materials and Application Technology of Ministry of Education, School of Materials Science and Engineering, Xiangtan University, Xiangtan 411105, China}
\affiliation{Hunan Provincial Key laboratory of Thin Film Materials and Devices, School of Materials Science and Engineering, Xiangtan University, Xiangtan 411105, China}
\author{Pan Zhou}
 \email{zhoupan71234@xtu.edu.cn}
 \affiliation{Key Laboratory of Low Dimensional Materials and Application Technology of Ministry of Education, School of Materials Science and Engineering, Xiangtan University, Xiangtan 411105, China}
\affiliation{Hunan Provincial Key laboratory of Thin Film Materials and Devices, School of Materials Science and Engineering, Xiangtan University, Xiangtan 411105, China}
\author{Baoru Pan}
 \affiliation{Key Laboratory of Low Dimensional Materials and Application Technology of Ministry of Education, School of Materials Science and Engineering, Xiangtan University, Xiangtan 411105, China}
 \affiliation{Hunan Provincial Key laboratory of Thin Film Materials and Devices, School of Materials Science and Engineering, Xiangtan University, Xiangtan 411105, China}
\author{Songmin Liu}
 \affiliation{Key Laboratory of Low Dimensional Materials and Application Technology of Ministry of Education, School of Materials Science and Engineering, Xiangtan University, Xiangtan 411105, China}
 \affiliation{Hunan Provincial Key laboratory of Thin Film Materials and Devices, School of Materials Science and Engineering, Xiangtan University, Xiangtan 411105, China}
\author{Binchang Zhou}
 \affiliation{Key Laboratory of Low Dimensional Materials and Application Technology of Ministry of Education, School of Materials Science and Engineering, Xiangtan University, Xiangtan 411105, China}
 \affiliation{Hunan Provincial Key laboratory of Thin Film Materials and Devices, School of Materials Science and Engineering, Xiangtan University, Xiangtan 411105, China}
\author{Lizhong Sun}
 \email{lzsun@xtu.edu.cn}
 \affiliation{Key Laboratory of Low Dimensional Materials and Application Technology of Ministry of Education, School of Materials Science and Engineering, Xiangtan University, Xiangtan 411105, China}
 \affiliation{Hunan Provincial Key laboratory of Thin Film Materials and Devices, School of Materials Science and Engineering, Xiangtan University, Xiangtan 411105, China}
\date{\today}
\begin{abstract}
 Research on altermagnetism has thus far primarily focused on spin-polarized bulk electronic states in magnetic materials. In this work, we advance the field by introducing the concept of surface altermagnetism (SAM), wherein altermagnetic spin polarization emerges at the surfaces of collinear antiferromagnets (AFMs) or altermagnets (AMs). To lay the theoretical groundwork for this phenomenon, we construct a thorough symmetry-based framework that systematically connects bulk spin groups to surface spin groups for both types of systems. Through symmetry analysis, we identify all symmetry-breaking surfaces capable of supporting SAM, identifying 35 for $PT$-symmetric AFMs and 61 distinct cases for bulk AMs. Moreover, we show that 203 collinear spin space groups---including 100 without and 103 with the $[C_2 \Vert P]$ operation---permit the appearance of SAM on the surface of $\bm{t}T$-symmetric AFMs via the breaking of fractional translational symmetries. The proposed framework is verified using tight-binding models and first-principles calculations, with practical material implementations shown in representative compounds like NaMnP, LiMnAs, and CrSb. Our results establish SAM as a robust, symmetry-protected magnetic state, extending altermagnetic phenomena to material surfaces and paving the way for advanced, field-free spin manipulation in next-generation spintronic technologies.
\\
\end{abstract}
\maketitle
\indent \emph{Introduction.}---The electronic and magnetic properties of crystalline materials are profoundly shaped not only by their bulk symmetries but also by the symmetries preserved or broken at their surfaces~\cite{Interf_mag}. Crystalline surfaces inherently reduce the symmetry of the parent bulk lattice, often giving rise to emergent electronic states~\cite{Tl,multi_ferro}, unconventional spin textures~\cite{spin_text_1,spin_text_2}, and exotic magnetic configurations~\cite{Skyrmion}. For instance, emergent surface multiferroicity appears on the surfaces of antiferromagnets (AFMs)~\cite{multi_ferro}, Rashba spin splitting at interfaces such as those in semiconductor heterostructures~\cite{spin_text_1}, and magnetically induced boundary phenomena like skyrmions at magnetic surfaces~\cite{Skyrmion}. These examples highlight a broader principle: the interplay between bulk and surface symmetries provides a fertile ground for uncovering novel states of matter.\\
\indent The recently discovered class of altermagnets (AMs) has sparked intense interest in condensed matter physics due to their unique nonrelativistic spin-split band structure without net magnetization~\cite{alter1,alter2,alter3,alter4,alter5,alter6,alter7,alter8,alter9,alter10,alter11,alter12,spin1,spin2,spin3,spin4}. Unlike conventional AFMs, where spin sublattices are related by translation or inversion symmetries, AMs exhibit a symmetry-driven spin splitting arising from the interplay of crystal rotation and magnetic order~\cite{alter1,alter2}. This leads to Fermi surface anisotropy and unconventional transport phenomena occurring without a macroscopic magnetic field~\cite{alter4,alter6}. Nevertheless, until now, theoretical and experimental research has concentrated on the bulk electronic states of these materials~\cite{alter7,MnTe_1,MnTe_2,CrSb_1,CrSb_2}, even though recent experiments indicate that surface states can significantly affect their transport properties and related characteristics~\cite{exp_RuO2}. This raises an intriguing question: could an analogue of altermagnetism emerge at crystal surfaces? Such boundary effects have been examined in individual systems~\cite{surf_alter1,surf_alter2,surf_alter3,surf_alter4,surf_alter5,surf_alter6,surf_alter7,surf_alter8}, and classified for strictly two-dimensional layers and for surface magnetic dipole order~\cite{surf_alter9,surf_alter10}, but a universal symmetry criterion for emergent altermagnetism at collinear magnetic boundaries, particularly for bulk spin-degenerate AFMs, remains lacking. \\
\indent In this paper, we introduce the surface altermagnetism (SAM) as a surface analogue of altermagnetism in collinear magnets, where symmetry reduction at the surface induces altermagnetic spin splitting. Crucially, SAM is not merely the surface manifestation of a bulk AM: it can emerge at the boundaries of conventional collinear AFMs that are spin degenerate in the bulk, identifying the boundary itself as the origin of the nonrelativistic spin splitting. To systematically elucidate its formation, we establish the group-subgroup symmetry relations between bulk and surface structures using spin group analysis, deriving all possible pathways. For collinear $PT$-symmetric AFMs and AMs, we identify 35 and 61 pathways within 21 and 37 spin point groups (SPGs), respectively, that enable SAM. Additionally, for $\bm{t}T$-symmetric AFMs, a total of 203 collinear spin space groups (SSGs; 100 without and 103 with the $[C_2 \Vert P]$ symmetry) permit transitions from $\bm{t}T$-symmetric AFM to SAM via the breaking of specific translational symmetries. To demonstrate the realization of SAM, we develop tight-binding models and perform first-principles calculations to some experimental synthesized three dimensional (3D) compounds such as NaMnP and LiMnAs (for $PT$- and $\bm{t}T$-symmetric AFMs) and CrSb (for AM).\\
\begin{figure}[t]
\centering
\includegraphics[trim={0.0in 0.0in 0.0in 0.0in},clip,width=\linewidth]{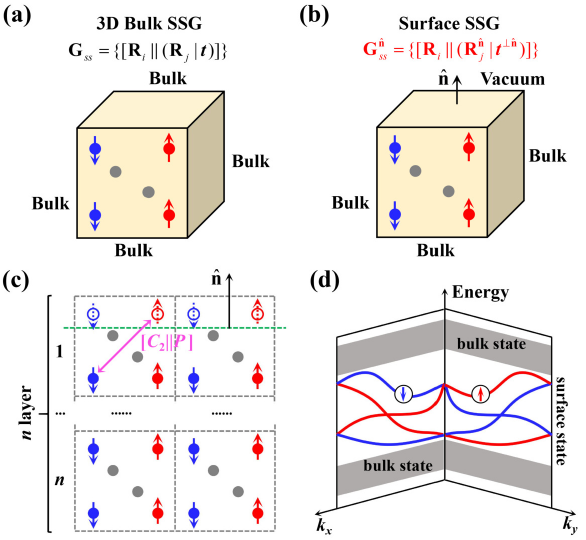}
\caption{Spin group representation of the SAM. (a) SSG description of a 3D collinear antiferromagnetic (or altermagnetic) material, with red, blue, and grey spheres denoting spin-up, spin-down, and nonmagnetic sites. (b) Surface SSGs for the 2D counterparts of semi-infinite surfaces. (c) Side view of the slab structure, where surface explicitly breaks $PT$ symmetry. (d) Surface states of slab structure, with grey indicating bulk states and red (blue) representing spin-up (spin-down) surface states.}\label{fig1}
\end{figure}
\indent \emph{Symmetry analysis.}---To realize the collinear SAM, the bulk materials must host two oppositely aligned spin sublattices, as illustrated in Fig. 1(a). Moreover, they must be connected by some symmetry operations, therefore, the bulk materials must be AFMs or AMs. In the nonrelativistic limit, the symmetries of these two kinds of systems are described spin groups, which can be expressed as $\textbf{r}_s$ $\times$ $\textbf{R}_{s}$\cite{sgroup1,sgroup2,sgroup3,sgroup4}. Here, $\textbf{r}_s$ represents the spin-only group and can be written as $\textbf{C}_\infty$ + $\bar{C}_2\textbf{C}_\infty$, a form common to all collinear spin groups, and $\textbf{R}_{s}$ refers to the second-type (AFM) or third-type (AM) nontrivial spin group:
\begin{equation}\label{eq1}
\textbf{G}_{ss} = \{[\textbf{R}_{i} \Vert (\textbf{R}_{j} \vert \bm{t})]\},
\end{equation}
where $\textbf{R}_{i} = E$ and $C_2$ denotes the identity and a twofold rotation about an axis perpendicular to the spin direction in spin space, respectively. The real-space operation $(\textbf{R}_{j} \vert \bm{t})$ includes spatial symmetries combined with a possibly fractional translation $\bm{t}$. When a crystal is cleaved along a given Miller plane $(hkl)$, a surface is created with vacuum on one side. The surface normal, represented by the unit vector $\hat{\textbf{n}}$, plays a central role in defining the resulting surface symmetry\cite{surfmag1,surfmag2}. According to the statement of Weber \textit{et al.}\cite{surfmag2}, the surface SSG corresponds to a subgroup of the bulk SSG [see Fig.~1(b)] and can be written as:
\begin{equation}\label{eq2}
\textbf{G}^{\hat{\textbf{n}}}_{ss} = \{[\textbf{R}_{i} \Vert (\textbf{R}^{\hat{\textbf{n}}}_{j} \vert \bm{t}^{\perp \hat{\textbf{n}}})]\},
\end{equation}
where only translational symmetries parallel to the surface are preserved, i.e., $\bm{t}^{\perp \hat{\textbf{n}}}$, while those with components along $\hat{\textbf{n}}$ are broken due to the reduced dimensionality at the surface. For a semi-infinite surface, spatial operations that reverse the surface normal are naturally broken, reducing the symmetry description to spin wallpaper groups [Tab.~S1 of the Supplementary Material (SM)]~\cite{sup}. As subperiodic spin groups~\cite{subperiodic1,subperiodic2,subperiodic3}, both spin wallpaper and layer groups are mapped to the bulk SSG in Appendix A and Secs.~I\,B and I\,C of the SM~\cite{sup}. \\
\indent Because SAM need two symmetry-related opposite sublattices, here we focus on surfaces where the two spin sublattices remain magnetically compensated, such that the net magnetization vanishes and no equilibrium surface magnetization is present. Since altermagnetic spin splitting primarily originates from SPG symmetries, our analysis is restricted to the surface SPG, denoted as $\textbf{g}^{\hat{\textbf{n}}} = \{[\textbf{R}_{i} \Vert \textbf{R}^{\hat{\textbf{n}}}_{j}]\}$. The SAM can emerge if this group lacks certain symmetry operations, such as $[C_2 \Vert C^{\hat{\textbf{n}}}_2] [\bar{C}_2 \Vert T]$ or $[C_2 \Vert \bm{t}^{\perp \hat{\textbf{n}}}]$, which would otherwise map one spin sublattice onto the other and suppress spin splitting on surface Brillouin zone (BZ). In fact, these two operations correspond to the surface versions of conventional $PT$ and $\bm{t}T$. Traditional collinear AFMs can generally be classified into two categories based on the symmetry that relates the opposite spin sublattices: $PT$-symmetric AFMs, where the connection is via the $[C_2 \Vert P] [\bar{C}_2 \Vert T]$ operation, and $\bm{t}T$-symmetric AFMs, where the sublattices are related by $[C_2 \Vert \bm{t}]$. In $PT$-symmetric AFMs, cleaving the crystal along certain surfaces---for instance, the plane marked by the green dashed line in Fig.~1(c)---can break the bulk $PT$ symmetry. If the operations of $[C_2 \Vert C^{\hat{\textbf{n}}}_2] [\bar{C}_2 \Vert T]$ or $[C_2 \Vert \bm{t}^{\perp \hat{\textbf{n}}}]$ also do not exist for the surface, the spin-up and spin-down surface states can exhibit the characteristic spin splitting of SAM, as shown in Fig. 1(d). For $\bm{t}T$-symmetric AFMs, SAM may emerge if all the aforementioned band-degeneracy-inducing operations are absent, and certain rotational or vertical mirror symmetries still connect the opposite spin sublattices. \\
\begin{figure}[t]
\centering
\includegraphics[trim={0.0in 0.0in 0.0in 0.0in},clip,width=\linewidth]{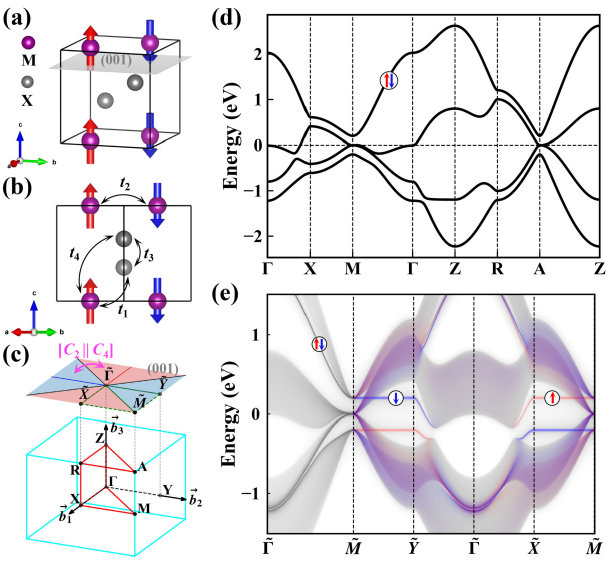}
\caption{Tight-binding model for a $PT$-symmetric AFM. (a) Lattice structure and (b) hopping configuration. Purple and gray spheres denote magnetic and nonmagnetic sites. (c) 3D and 2D projected BZs for the tetragonal lattice. (d) Bulk energy band structure, calculated with $t_1 = 0.4$, $t_{2} = 0.3$, $t_{3} = 0.2$, $t_4 = -0.15$, and $J = 0.2$. (e) Surface states for the (001) surface, with gray for bulk states and black, red, and blue for spin-degenerate, spin-up, and spin-down states, respectively.}\label{fig2}
\end{figure}
\indent Based on group-theoretical analysis, the SAM can be systematically understood by examining bulk-to-surface symmetry relations within the framework of group-subgroup structures, and it can be classified into three distinct types: $PT$-symmetric antiferromagnetic SAM ($PT$-AFM-SAM), $\bm{t}T$-symmetric antiferromagnetic SAM ($\bm{t}T$-AFM-SAM), and altermagnetic SAM (AM-SAM). For $PT$-AFM-SAM and AM-SAM, both are characterized by SPGs. We systematically analyze all 58 collinear SPGs and their corresponding subgroups, identifying 61 symmetry-allowed pathways for AM-SAM and 35 for $PT$-AFM-SAM. These pathways are comprehensively documented in Tabs. S3 and S4 of the SM~\cite{sup}. For the $\bm{t}T$-AFM-SAM class, we investigate 517 collinear SSGs associated with collinear $\bm{t}T$-symmetric AFMs. Among these, 203 SSGs are found to support SAM when cleaved along specific surfaces, as detailed in Tab. S5 of the SM~\cite{sup}. In total, our analysis identifies 61, 35, and 203 symmetry-allowed cases for AM-SAM, $PT$-AFM-SAM, and $\bm{t}T$-AFM-SAM, respectively. A complete overview of these results is provided in Tabs. S1--S5 of the SM~\cite{sup}. Although the bulk operation governing opposite-spin degeneracy dictates whether SPG or SSG applies, all compensated semi-infinite surfaces ultimately unify into spin wallpaper groups under eleven SPGs (Tab.~S1, Appendix A, and Sec.~I\,D of the SM)~\cite{sup}.  \\
\indent \emph{Tight-binding model.}---To elucidate the formation of the SAM, we develop a tight-binding model that captures the electronic structure of a collinear AFM. This model is based on a crystal structure with the SPG ${^{2}4}/{^{2}m}{^{1}m}{^{2}m}$, which exhibits $PT$ symmetry, as illustrated in Fig. 2(a). The structure comprises two magnetic atoms M at the Wyckoff position 2a and two nonmagnetic atoms X at the Wyckoff position 2c. Using this framework, we construct the Hamiltonian for the $PT$-symmetric antiferromagnetic system as follows:
\begin{equation}\label{eq3}
\begin{aligned}
\mathcal{H}_{AFM} & = t_{1}\sum_{i,j}c_{\alpha,i}^{\dagger}c_{\beta,j} + t_{2}\sum_{i,j}c_{\beta,i}^{\dagger}c_{\beta,j}  + t_{3}\sum_{i,j}c_{\alpha,i}^{\dagger}c_{\alpha,j} \\ &
+ t_{4}\sum_{i,j}c_{\alpha,i}^{\dagger}c_{\beta,j} + J\sum_{\beta,i}c_{\beta,i}^{\dagger}\tau_{z}\sigma_{z}c_{\beta,i} + h.c.,
\end{aligned}
\end{equation}
where $c_{\alpha(\beta),i}^\dagger$ and $c_{\alpha(\beta),i}$ represent the creation and annihilation operators for electrons on sublattice $\alpha$ (nonmagnetic) and $\beta$ (magnetic) at site $i$, respectively. A comprehensive explanation of the model can be found in Appendix B.\\
\indent By cleaving the (001) surface along the shaded plane perpendicular to the $c$-axis, as depicted in Fig. 2(a), the resulting surface consists of terminations formed by nonmagnetic atoms. This cleavage disrupts the $PT$ symmetry of the system. However, the surface retains specific symmetries, including the $[C_{2}\Vert{C_{4}}]$ and the vertical mirror operations $[C_{2}\Vert{M_{110}}]$ and $[C_{2} \Vert M_{\bar{1}10}]$. The symmetry reduction resulting from the surface cleavage lowers the SPG to $^{2}4^{1}m^{2}m$. As shown in Fig. 2(e), this reduced surface SPG causes the surface states to exhibit momentum-dependent alternating spin splitting along the $\tilde{M}-\tilde{Y}-\tilde{\Gamma}-\tilde{X}-\tilde{M}$ path. When examining the spin distribution across the entire BZ, these surface states reveal a characteristic $d$-wave distribution, as depicted in Fig. 2(c). For both $\bm{t}T$-symmetric antiferromagnetic and altermagnetic systems, similar results are obtained from their corresponding tight-binding models. The SAM can emerge as long as the surface breaks both $[C_2 \Vert P] [\bar{C}_2 \Vert T]$ and $[C_2 \Vert \bm{t}]$ symmetries, while preserving certain symmetries that connect opposite spin sublattices. Detailed model constructions and results are provided in Sec. II of the SM~\cite{sup}. We stress that the AM case is not a mere transcription of the bulk: because the surface symmetry of an AM is generally lower than that of its bulk, SAM in AMs need not reproduce the bulk spin splitting, and the surface and bulk states can differ physically.\\
\begin{figure}[b]
	\centering
	\includegraphics[trim={0.0in 0.0in 0.0in 0.0in},clip,width=\linewidth]{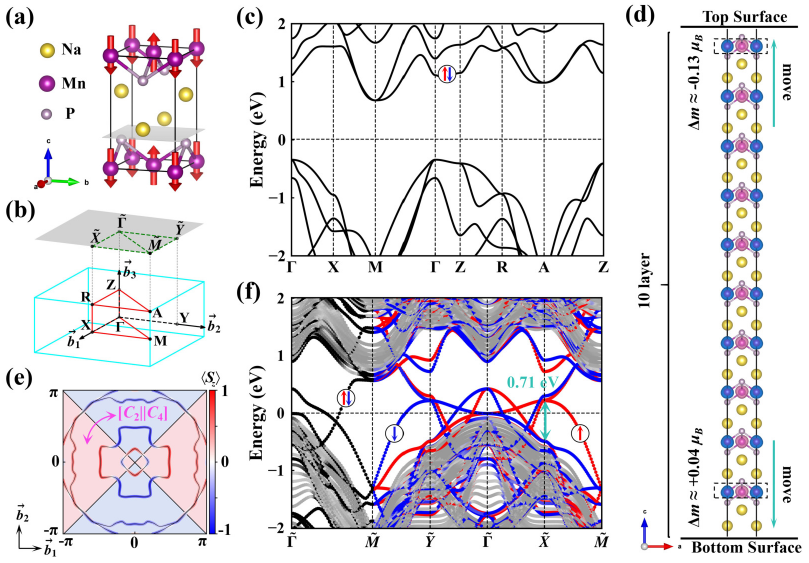}
	\caption{Realization of the SAM in $PT$-symmetric NaMnP. (a) Crystal structure, (b) 3D BZ, and (c) bulk energy band structure. (d) Surface relaxation of 10-layer NaMnP slab. (e) Spin-resolved Fermi surface at $E$ = 0.15 eV, showing $d$-wave-like symmetry on the (001) surface. Color map indicates out-of-plane spin component $\langle S_z \rangle$. (f) Slab band structure for the (001) surface, showing altermagnetic spin splitting. Gray denotes bulk states, while black, red, and blue represent spin-degenerate, spin-up, and spin-down states, respectively.}\label{fig3}
\end{figure}
\indent \emph{Realization in realistic materials.}---Here, we demonstrate the realization of the SAM in collinear antiferromagnetic and altermagnetic systems through first-principles calculations, and the detailed computational methods are provided in the Sec. III of the SM\cite{sup}. For $PT$-symmetric AFMs, we select NaMnP~\cite{NaMnX_1,NaMnX_2,NaMnX_3} as a representative candidate, whose crystal structure is depicted in Fig. 3(a). This material exhibits antiferromagnetic ordering at room temperature, with a N\'eel temperature of 560 K~\cite{NaMnX_1}. Its SPG is ${^{2}4}/{^{2}m}{^{1}m}{^{2}m}$, which aligns precisely with the symmetry of our theoretical $PT$-symmetric AFM model. The electronic structure calculations, carried out using the GGA+$U$ approach within the Dudarev formalism with an effective Hubbard parameter $U_\mathrm{eff}$ = 3 eV applied to Mn atoms, reveal semiconducting behavior with a band gap of 1.02 eV, as illustrated in Figs. 3(b) and 3(c). Due to the preserved $PT$ symmetry in the bulk, the spin-up and spin-down electronic states remain fully degenerate across the entire BZ.\\
\indent For NaMnP, the (001) surface admits three inequivalent terminations, named here by their top-bottom atomic layers. We focus on the P-Na terminated one, which, when cleaved orthogonally to the $c$-axis, lacks inversion symmetry, breaking the bulk $[C_2\Vert P][\bar{C}_2\Vert T]$ symmetry but preserving $C_4$ rotation and vertical mirror symmetries [see Figs. 3(a) and 3(d)]. Structural relaxation shows no surface reconstruction, only slight atomic displacements toward the surface [Fig. 3(d)], and the surface magnetic ground state coincides with the bulk order, so that the slab symmetry remains $^{2}4^{1}m^{2}m$. The preserved $C_4$ and mirror symmetries ($M_{110}$ and $M_{\bar{1}10}$) between opposite spin sublattices lead to alternating spin polarization across the mirror planes, resulting in a $d$-wave-like spin-splitting pattern, as shown in Figs. 3(e) and 3(f). The three terminations, the competing surface magnetic configurations, and the influence of spin-orbit coupling (SOC) are analyzed in Appendix C and Sec. IV of the SM~\cite{sup}. \\
\indent Beyond NaMnP, three further $PT$-symmetric AFMs, CaMnBi$_2$~\cite{CaMnBi2_1}, KFeS$_2$~\cite{KFeS2_1,KFeS2_2,KFeS2_3}, and CaMn$_2$O$_4$~\cite{CaMn2O4}, display altermagnetic spin splitting at their surfaces (Sec. IV of the SM\cite{sup}). The $\bm{t}T$-symmetric AFMs form a second important class of conventional AFMs, for LiMnAs~\cite{LiMnAs}, SrRu$_2$O$_6$~\cite{SrRu2O6_1,SrRu2O6_2}, and MnS$_2$~\cite{MnS2_1,MnS2_2}, we likewise obtain spin-split surface states (Sec. V of the SM\cite{sup}). Altogether, first-principles calculations for ten synthesized compounds across all three SAM classes (Secs. IV--VI of the SM~\cite{sup}) corroborate the classification, which Tabs. S1--S5~\cite{sup} render predictive: the bulk SSG of any collinear magnet directly yields the surface orientations capable of hosting SAM.  \\
\begin{figure}[t]
\centering
\includegraphics[trim={0.0in 0.0in 0.0in 0.0in},clip,width=\linewidth]{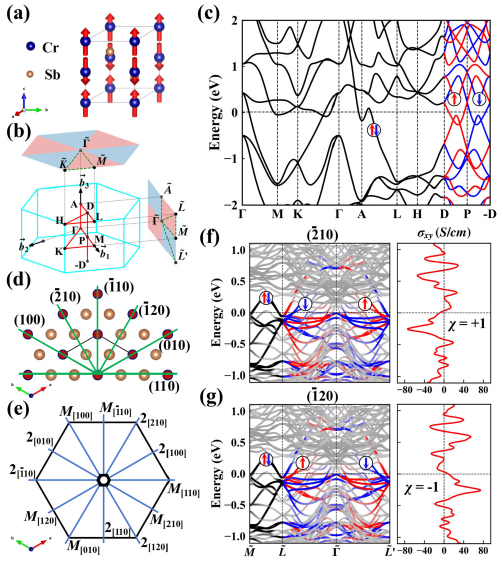}
\caption{Realization of SAM in altermagnetic CrSb. (a) Crystal structure. (b) 3D BZ with top cross-section (altermagnetic spin splitting) and projection on the $(\bar{2}10)$ and $(\bar{1}20)$ planes. (c) Bulk electronic bands. (d, e) Surface terminations with different crystallographic orientations and their symmetry operations. (f, g) Slab band structures and anomalous Hall conductivity for $(\bar{2}10)$ and $(\bar{1}20)$ surfaces. Bulk, spin-degenerate, spin-up, and spin-down states are colored gray, black, red, and blue, respectively.}\label{fig4}
\end{figure}
\indent Next, we discuss the SAM of AMs using CrSb~\cite{CrSb_3,CrSb_4,CrSb_5}, a $g$-wave altermagnetic material (further AM examples, RuO$_2$~\cite{RuO2_obser,RuO2_model} and MnSeO$_3$~\cite{MnSeO3_1,MnSeO3_2}, are treated in Sec. VI of the SM~\cite{sup}) with SPG ${^{2}{6}}/{^{2}m}{^{1}m}{^{2}m}$ [Fig. 4(a)]. CrSb, with a high N\'eel temperature (above 700 K) and significant spin splitting [approximately 1.2 eV near the Fermi level\cite{CrSb_1}, Figs. 4(b)-4(c)], is a promising spintronic candidate. In the bulk, opposite-spin sublattices are related by composite symmetry operations like $[C_{2} \Vert (C_{6} \vert 00\frac{1}{2})]$ and $[C_{2} \Vert (M_{001} \vert 00\frac{1}{2})]$, leading to spin-degenerate nodal planes. Cleaving along certain crystal surfaces breaks these fractional translations, reducing the spin group to a subgroup. For example, the (001) surface breaks the fractional translation $(0, 0, \frac{1}{2})$, reducing the spin group to its ferromagnetic subgroup $^{1}3^{1}m$, causing spin splitting across the surface BZ, as shown in Fig. S13(b)\cite{sup}. \\
\indent Beyond the (001) surface, two surface groups parallel to the $c$-axis were examined: \{(100), (010), ($\bar{1}$10)\} and \{($\bar{2}$10), ($\bar{1}$20), (110)\}, as illustrated in Fig. 4(d). The first group preserves $M_{001}$ mirror symmetry, $C^{\hat{\textbf{n}}}_{2}$ rotation symmetry, and $M^{\parallel\hat{\textbf{n}}}$ vertical mirror symmetry, reducing the SPG to ${^{2}m}{^{1}m}{^{2}2}$, and the surface band structure remains fully spin-degenerate [Fig. S13(c)]\cite{sup}. The second group also preserves $C^{\hat{\textbf{n}}}_{2}$, forming the symmetry subgroup ${^{2}m}{^{2}m}{^{1}2}$. For the ($\bar{2}$10) surface, the surface normal $\hat{\textbf{n}}$ aligns with $[100]$, and in-plane vectors $\vec{\bm{b}}_1$ and $\vec{\bm{b}}_2$ align with $[001]$ and $[120]$ [see Fig. S14(a)]. Here, the $C_{2_{100}}$ ($C^{\hat{\textbf{n}}}_{2}$) rotation symmetry connects identical spin states, while the mirror symmetries $M_{001}$ and $M_{120}$ link opposite spin states. As shown in Fig. 4(f), the ($\bar{2}$10) surface band structure is spin-degenerate along the surface basis vectors, but exhibits spin splitting along the $\tilde{L}-\tilde{\Gamma}-\tilde{L}'$ diagonal, manifesting $d$-wave characteristics. A similar spin splitting is seen for the ($\bar{1}$20) surface [see Fig. 4(g)], but with reversed spin polarization, highlighting the potential for spin-polarization dependent engineering of spintronic devices through selective surface termination (see Appendix D and Sec. VII of the SM~\cite{sup} for details). \\
\indent \emph{Discussion.}---In conclusion, we have developed a theoretical framework and implementation strategy for inducing SAM, providing a controllable and widely applicable route to altermagnetism on specific crystallographic surfaces of AFMs or AMs. Our work is distinguished by three key attributes: (1) \textbf{Universality}: The strategy is applicable to a broad range of conventional collinear magnets, encompassing AMs, $PT$- and $\bm{t}T$-symmetric AFMs. (2) \textbf{Tunability}: Surface states can be tailored by surface orientation and atomic composition, enabling intrinsic control over the spin states. (3) \textbf{Surface-specific responses}: SAM activates observables that are absent in the corresponding bulk. This is exemplified by the surface crystal Hall effect of CrSb, which is symmetry-forbidden in the bulk yet emerges with a chirality-controlled sign on the ($\bar{2}$10) and ($\bar{1}$20) facets, and by spin-splitter-type transport at the boundaries of AFMs whose bulk bands are spin degenerate. \\
\indent The SAM proposed here can be detected by several techniques. In many conventional AFMs the bulk anomalous Hall conductivity vanishes by symmetry, whereas the symmetry-broken surfaces of SAM may host a surface anomalous Hall effect, offering an indirect probe of surface altermagnetic phases~\cite{surf_ano1,surf_ano2,surf_ano3}. Furthermore, leveraging recent advancements in experimental studies of AMs, spin- and angle-resolved photoemission spectroscopy (spin-ARPES) provides a powerful tool to directly probe the band dispersion of both surface and bulk electronic states, revealing their distinct spin and momentum characteristics~\cite{arpes1,arpes2,arpes3}. The embedding depth of SAM can also be probed experimentally (Sec. IV of the SM~\cite{sup}), and spin-polarized scanning tunneling microscopy (STM) and magneto-optical Kerr effect (MOKE) measurements can further validate the surface spin textures. The magnitude of the splitting and its detectability against parallel bulk conduction are addressed in Appendix E. 

\indent \emph{Acknowledgments.}---This work is supported by the National Natural Science Foundation of China (Grant No. 12574070 and No. 12504223), the Postgraduate Scientific Research Innovation Project of Hunan Province (CX20240616), the China Postdoctoral Science Foundation(Grants No. 2025M773383 and No. GZC20252231), the China Postdoctoral Science Foundation-Hunan Joint Support Program(Grant No. 2025T002HN). 

\indent Yuzhong Hu and Pan Zhou contributed equally to this work.

\indent \textit{Note added.}---After the submission of this manuscript, we became aware of a related independent work that presents a full symmetry classification and material identification of emergent altermagnetism at surfaces of antiferromagnets~\cite{note_added1}.
\bibliography{references}
\onecolumngrid
\vspace{1em}
\begin{center}
	\textbf{\large End Matter}
\end{center}
\vspace{1em}
\twocolumngrid

\indent \textit{Appendix A: Surface symmetry groups and the levels of the classification}---The surface groups are subperiodic spin groups: cleaving the bulk to a finite slab yields a spin layer group (a 2D translation lattice with 3D point operations), and the semi-infinite surface further removes every operation reversing $\hat{\textbf{n}}$, leaving a spin wallpaper group whose spatial parent is one of the ten 2D crystal classes. Enumerating their halving subgroups gives exactly eleven admissible surface SPGs (Tab. S1 of the SM~\cite{sup}), consistent with Ref.~\cite{note_added1}. A wallpaper-group description of surface magnetic symmetry is thus contained in the present construction, whose additional content is the group-subgroup map from the bulk SSG. \\
\indent The three SAM classes differ only in their bulk parentage: whatever the parent phase, the reduction terminates in one of eleven surface SPGs of Tab. S1~\cite{sup}, so that the surface description is uniform. What differs is which bulk symmetry has to be removed by the termination, and hence how much of the bulk group must be examined. An element $[\mathbf{R}_i \Vert (\mathbf{R}_j \vert \bm{t})]$ survives at a surface with normal $\hat{\textbf{n}}$ only if $\mathbf{R}_j$ leaves $\hat{\textbf{n}}$ invariant and $\bm{t}$ has no component along it. In $PT$-symmetric AFMs the spin degeneracy of the two sublattices is enforced by $[C_2 \Vert P][\bar{C}_2 \Vert T]$, whereas in AMs, whose bulk bands are already spin split, the sublattices are connected by rotations or rotoinversions. In both cases the operations concerned carry a nontrivial point part, so that the surviving group is again labelled by a surface SPG, and enumerating the 58 collinear SPGs with their orientation-resolved subgroups is complete and nonredundant. In $\bm{t}T$-symmetric AFMs, by contrast, the sublattices are connected solely by a fractional translation $[C_2 \Vert \bm{t}]$, whose point-group image is the identity: an SPG-level treatment would render it as a spin flip with trivial spatial action and would therefore be blind to the very symmetry whose removal generates SAM. Whether $[C_2 \Vert \bm{t}]$ survives depends on the component of $\bm{t}$ along $\hat{\textbf{n}}$, i.e., on the interplay between the translation part of the SSG and the Miller indices. This class must therefore be analyzed at the SSG level, which is why the 517 collinear SSGs of $\bm{t}T$-symmetric AFMs were examined individually. The full construction is given in Secs. I\,B--I\,D of the SM~\cite{sup}, where the bulk-to-slab and slab-to-surface reductions are derived in turn, the eleven admissible surface SPGs are listed in Tab. S1, and the resulting pathways are enumerated in Tabs. S3--S5. \\
\indent It is worth emphasizing what this construction classifies. The object is not a set of 2D spin groups, which would duplicate existing collinear classifications, but the pair consisting of a bulk SSG and a surface orientation $\hat{\textbf{n}}$, together with the group--subgroup reduction it induces. Each such pair therefore constitutes one symmetry-allowed pathway, which is the object enumerated in the classification, and the three SAM classes are labelled by which bulk sublattice-connecting symmetry the termination removes: $[C_2 \Vert P][\bar{C}_2 \Vert T]$ for $PT$-AFM-SAM, the fractional translation $[C_2 \Vert \bm{t}]$ for $\bm{t}T$-AFM-SAM, and rotations or rotoinversions for AM-SAM. This information is absent from any bulk classification, and it is what makes the scheme orientation resolved. CrSb illustrates the point: one and the same bulk spin group yields an uncompensated, ferromagnetic-subgroup surface on (001), fully spin-degenerate surfaces on $\{(100),(010),(\bar{1}10)\}$, and $d$-wave SAM with facet-dependent sign of the spin polarization on $\{(\bar{2}10),(\bar{1}20),(110)\}$---precisely the orientation-resolved information required by surface-sensitive experiments. \\
\indent The classification treats spin and spatial symmetries independently, deriving the surface groups from the bulk SSG through symmetry reduction at the termination. For long-range magnetic order and negligible SOC, it is closed, self-consistent, and offers an exhaustive symmetry-based description of SAM\cite{sgroup1,sgroup2,sgroup3,sgroup4}, although its applicability may be limited by strong SOC, significant surface reconstruction, disorder, or correlation-driven symmetry breaking. We also emphasize that the SAM proposed here is fundamentally distinct from the Rashba effect originating from relativistic SOC. Further discussions regarding its applicability and the distinction from the Rashba effect are provided in Secs. I\,E and I\,F of the SM~\cite{sup}.

\indent \textit{Appendix B: $PT$-symmetric antiferromagnetic model}---Here, we present a detailed description of the $PT$-symmetric antiferromagnetic model, where the matrices $\tau_i$ and $\sigma_i$ are Pauli matrices acting on the sublattice and spin degrees of freedom, respectively. The first and fourth terms capture the nearest-neighbor and next-nearest-neighbor hopping (involving parameter $t_1$ and $t_4$) between magnetic and nonmagnetic atoms in Fig. 2(b), respectively. The second (third) term with parameter $t_2$ ($t_3$) describes hopping between magnetic (nonmagnetic) sites. The last term including parameter $J$ counts the AFM ordering with an out-of-plane easy axis. In this model, the $[C_2 \Vert P] [\bar{C}_2 \Vert T]$ symmetry ensures that the bulk band structure remains fully spin-degenerate, as illustrated in Figs. 2(c) and 2(d).

\indent \textit{Appendix C: Surface terminations and the validity domain of SAM}---For NaMnP, the (001) surface has three possible terminations (see Fig. S3): Mn-P, P-Na, and Na-Na (denoted by top-bottom atomic layers). To determine the magnetic ground state of the P-Na termination discussed in the main text, slab calculations with magnetic rearrangements in the outermost layers were conducted. As shown in Fig. S4, six possible surface magnetic configurations were considered, and energy comparisons indicate the lowest energy occurs when the magnetic order aligns with the bulk phase. Further details on surface terminations, magnetic configurations, and the influence of SOC on the SAM of NaMnP are discussed in the Sec. IV of the SM\cite{sup}.\\
\indent It should be emphasized that, in real materials, surface atomic and magnetic configurations can vary with cleavage plane, relaxation, or experimental conditions. Our calculations focus on terminations that preserve the antiferromagnetic sublattice structure, ensuring that the opposite-spin sublattices remain well-defined and related by spin symmetry. This condition is crucial for the emergence of SAM. Terminations that break sublattice equivalence or spin symmetry are expected to lift degeneracy protection and alter the spin-split surface bands. Thus, the reported SAM applies only to terminations preserving the antiferromagnetic sublattice symmetry. The present results therefore describe ideal terminations and identify promising surface orientations. Importantly, surface reconstruction does not invalidate the framework itself, since once the reconstructed atomic and magnetic structure is determined, experimentally or from first principles, the same group-theoretical procedure applies directly to its actual surface symmetry.

\indent \textit{Appendix D: Physical effects of SAM}---The observation that different surfaces of CrSb display opposite spin polarization suggests that surface termination offers a direct and effective way to control the spin polarization in SAM. A natural consequence is a surface-dependent transport response, which can be captured by the crystal Hall effect (CHE) as a direct manifestation of crystal chirality. In the bulk, CrSb belongs to the magnetic point group $6'/m'm'm$, for which the CHE is symmetry forbidden~\cite{alter4,AHE_1}. In contrast, we explicitly show that SAM activates a Hall response at the surface of CrSb. The two surface terminations, ($\bar{2}$10) and ($\bar{1}$20), belong to the magnetic point group $m'm'2$, and they possess opposite surface chiralities, which leads to a sign reversal of the anomalous Hall conductivity, as shown in Figs. 4(f)-4(g) and Fig. S19. This behavior is directly analogous to the chirality-controlled Hall effect in bulk collinear AMs and provides a clear transport signature of SAM.\\ 
\indent The bulk-to-boundary construction developed here is not restricted to vacuum-terminated surfaces, but carries over unchanged to interfaces, heterostructures, and electrically controlled boundary layers. Once SOC is incorporated, so that the spin-group description is replaced by the corresponding magnetic subperiodic groups, the same symmetry reduction applies to buried boundaries and provides a systematic route for analyzing magnetic interfaces, interface-induced magnetocrystalline anisotropy, and spin-dependent tunneling in vertical device geometries such as magnetic tunnel junctions, where interfacial anisotropy is a key performance parameter. Replacing the vacuum by a gate dielectric likewise leaves the symmetry analysis unchanged, so that gate-induced accumulation or inversion layers at the boundaries of magnetic semiconductors can be examined in the same manner, with the band bending providing an additional electrical handle on the spin-split boundary states.

\indent \textit{Appendix E: Physical strength and experimental detectability of SAM}---While the universality of SAM is symmetry-dictated, its magnitude is material- and surface-dependent. In NaMnP, first-principles calculations reveal a significant surface altermagnetic spin splitting of approximately 0.71~eV [see Fig. 3(f)], surpassing typical spin splittings (dominated by relativistic effects or ferromagnetic exchange) in many spintronic systems, and making SAM detectable by current surface-sensitive techniques. The momentum-dependent surface spin polarization can be probed using spin-ARPES. SAM may also produce characteristic transport signatures, such as direction-dependent spin-polarized currents or spin-splitting-induced anisotropies, detectable through nonlocal spin transport or spin Hall-like measurements, without the need for ferromagnetic contacts. These results demonstrate that SAM in NaMnP is symmetry-allowed and exhibits spin splitting that is sufficiently strong to be experimentally detectable. Further details is provided in Sec. IV of the SM~\cite{sup}.\\
\indent Because the spin splitting associated with SAM is confined to the uppermost surface layers, the surface conduction channel is connected in parallel with the bulk one. In samples thick enough to be regarded as bulk, conventional bulk-dominated spin-transport measurements may therefore have limited sensitivity to it, particularly when the bulk is metallic. Spin-transport detection is expected to be most suitable for thin films with controlled terminations, surface-dominated conducting channels, and interface-sensitive device geometries. Insulating or semiconducting AFMs realize such a surface-dominated channel intrinsically: in NaMnP, whose bulk gap is 1.02~eV, the in-gap conduction is carried by the spin-split surface states, so that the surface response is not shorted by the bulk. For thick samples, direct surface-sensitive probes, spin-ARPES and spin-polarized STM, provide the more direct means of examining the predicted surface states, with nonlocal geometries offering a complementary route.
\end{document}